\documentclass[12pt,notitlepage]{article}
\usepackage{eucal,amsmath,amsfonts,amsthm,amssymb,latexsym,epsfig}

\theoremstyle{plain}

\newtheorem{thm}{Theorem}

\newtheorem{Lma}{Lemma}
\newtheorem{Crly}{Corollary}

\theoremstyle{definition}

\newcommand{\rd}{\mathrm{d}}
\newcommand{\rw}{\mathrm{w}}

\newcommand{\ww}{\mathsf{\omega}}

\newcommand{\lp}{\overline{\lim}}
\newcommand{\lf}{\underline{\lim}}

\newcommand{\lindent}
{
\setlength{\labelwidth}{2cm}\setlength{\leftmargin}{2.0cm}
\setlength{\labelsep}{0.5cm}\setlength{\rightmargin}{1.0cm}
\setlength{\parsep}{0.5ex plus 0.2ex minus0.1ex}
\setlength{\itemsep}{0ex plus 0.2ex}
}

\newcounter{prop}

\newcommand{\ms}{{\setcounter{equation}{0}}{\setcounter{Lma}{0}}{\setcounter{Crly}{0}}\section}
\begin{document}
\setcounter{page}{1}

\title{Far Field Behavior of Noncompact Static Spherically Symmetric solutions of Einstein SU(2) Yang Mills Equations \thanks{Research performed while a Visiting Assistant Professor at Brown University}}
\author{Alexander N. Linden\thanks{Zorn Assistant Professor, Indiana University}}
\date{}
\maketitle


\begin{abstract}
There exist static spherically symmetric solutions of the Einstein equations with cosmological constant $\Lambda$ coupled to the SU(2) Yang Mills equations that are smooth at the origin $r=0$ and with an horizon which can be transformed away with a change of coordinates in which the radius increases across the singularity.  We establish the global behavior of these solutions.

\end{abstract}
\ms{Introduction}
\label{global}
$\;$

\cite{L2} analyzes spherically symmetric static solutions of the Einstein~-~SU(2) Yang Mills equations with small positive cosmological constant that are smooth at the origin of spherical symmetry.  In particular, it is proved that the presence of a positive cosmological constant $\Lambda$ causes each such solution to give rise to an horizon at some $r_c\le\sqrt{3/\Lambda}$.  For small $\Lambda$ a class of solutions was found in which this singularity is only due to choice of Schwarzschild coordinates and can be transformed away with a Kruskal-like change of coordinates in such a way that $r$ increases in the extended solution.  Furthermore, the Yang Mills curvature is well behaved under the change of coordinates.  In this paper we prove that such solutions are defined globally in Schwarzschild coordinates and that that the solutions (except for the one coordinate singularity) are everywhere smooth.  We also determine their asymptotic behavior.

With a spherical symmetric metric
\begin{equation}\label{metric}
\rd s^2=C^2A\;\rd t^2 -\frac{1}{A}\; \rd r^2 - r^2\;\rd \Omega^2
\end{equation}
($\rd \Omega^2=\rd \phi^2+\sin^2\phi\;\rd \theta^2$, the standard metric on the unit 2-sphere) and 
spherically symmetric connection on an $SU(2)$ bundle
\begin{eqnarray}\label{connection}
\ww&=& \mathrm{a}(r,t) \mathbf{\tau}_3 \;\rd t + \mathrm{b}(r,t)\mathbf{\tau}_3 \;\rd r +\rw(r,t)\mathbf{\tau}_2
\;\rd \phi\nonumber\\
&+&(\cos\phi\mathbf{\tau}_3-\rw(r,t) \sin \phi
\mathbf{\tau}_1)\;\rd \theta.
\end{eqnarray}
the Einstein~-~Yang Mills equations take the form of three ordinary differential equations for the coefficients $A$, $C$, and $\rw$,
\begin{equation}\label{Aeq}
rA'+2A{\rw'}^2=1-A-\frac{(1-\rw^2)^2}{r^2}-\Lambda r^2,
\end{equation}
\begin{equation}\label{weq}
r^2A{\rw}''+r(1-A-\frac{(1-\rw^2)^2}{r^2}-\Lambda r^2)+\rw(1-\rw^2)=0,
\end{equation}
and
\begin{equation}\label{Ceq}
rC'=2{\rw'}^2C.
\end{equation}
The $\tau_i$ in equation~(\ref{connection}) are the following matrices which form a basis of $su(2)$:
\[\mathbf{\tau}_1=i/2\left[
\begin{array}{cc}
0 & -1\\
-1 & 0
\end{array}
\right],\;
\mathbf{\tau}_2=i/2\left[
\begin{array}{cc}
0 & i\\
-i & 0
\end{array}
\right],\;
\mathbf{\tau}_3=i/2\left[
\begin{array}{cc}
-1 & 0\\
 0 & 1
\end{array}
\right].
\]
By appropriate change of gauge, one of $a$ and $b$ can be made to vanish.  We asume that in this gauge, the other vanishes too.

There are known explicit solutions to equations~(\ref{Aeq})~-~(\ref{Ceq}).  Among these is the Schwarzchild~-~deSitter metric with constant Yang Mills connection
\begin{equation}\label{deSitter}
\rd s^2 = (1-\frac{\Lambda r^2}{3})\;\rd t^2-(1-\frac{\Lambda r^2}{3})^{-1}\;\rd r^2 - r^2\;\rd\Omega^2,\;\;\;\rw\equiv 1.
\end{equation}
This solution is defined is defined for all $r$ and is smooth except at the horizon which occurs on the sphere $r=r_h=\sqrt{3/\Lambda}$ and serves as a prototype of solutions which have the following characteristic:
\begin{equation}\label{w'fin}
\lim_{r\nearrow r_h}\rw^2(r)<\infty\;\mathrm{and}\;\lim_{r\nearrow r_h}A'(r)<0.
\end{equation}
We call solutions that satisfy equation~(\ref{w'fin}) \emph{noncompact}.  For such solutions, a change in coordinates $(t,r)$ to new coordinates $(u,v)$ can be found that transforms the metric~(\ref{metric}) to
\begin{equation}\label{kmetric}
\rd s^2=g(u,v)(\rd u^2-\rd v^2)-r^2(u,v)\;\rd \Omega^2
\end{equation}
such that $g\not=0$ in a neighborhood of the singularity at $r_h$ \cite{rA75},\cite{jS951}. Constant $r$ and $t$ curves are shown in Figure~1.\\
\begin{figure}\label{prdiag}
\epsfig{figure=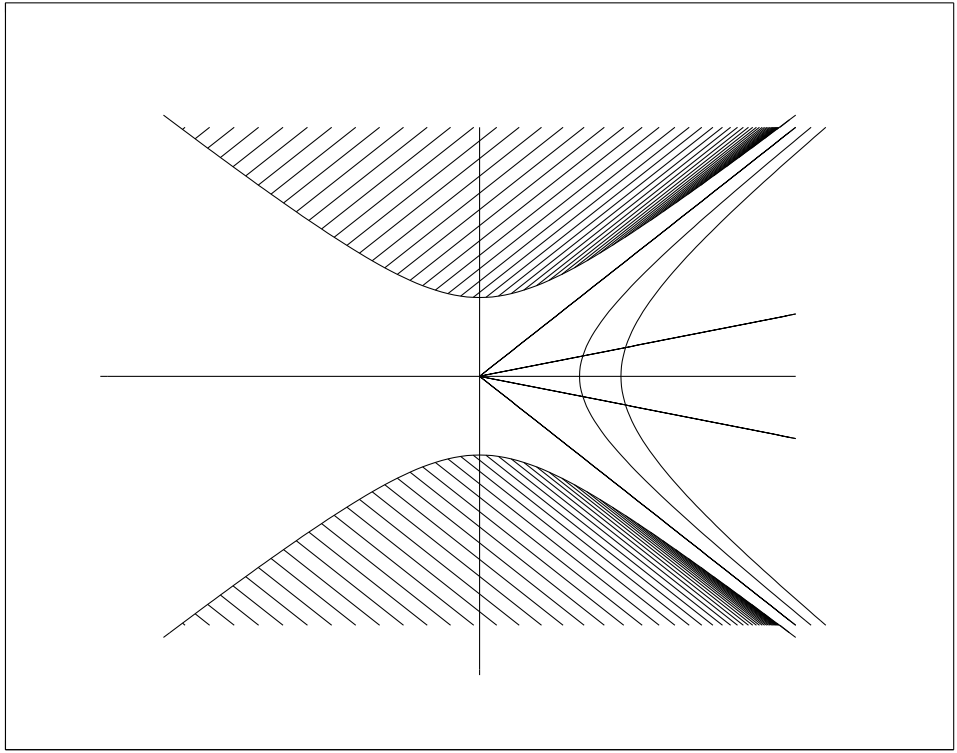,scale=.8}
\caption{\textbf{Spacetime geometry near r}$\mathbf{=r_h}$.  The spacetime manifold $M$ is the unshaded region.  The hyperbolas are curves of constant $r$ and the rays through the origin are curves of constant $t$.  The rays at angles $\pm 45^\circ$ represent $t=\pm\infty$ respectively.  The hyperbola for $r=r_h$ degenerates to the point at the origin.}
\end{figure}
\bigskip

\ms{Global Behavior}
We now prove that any solution of equations~(\ref{Aeq})~-~(\ref{Ceq}) that satisfies equation~(\ref{w'fin}) behaves qualitatively like the solution~(\ref{deSitter}); namely, the solution is defined and smooth for all $r\ne r_h$, the metric approaches the metric (\ref{deSitter}) as $r\nearrow \infty$, and the Yang Mills connection approaches a finite $\rw_\infty$ asymptotically as $r\nearrow \infty$.

Because the singularity at $r_h$ of such a solution is only a coordinate singularity and because of condition~(\ref{w'fin}), the solution $(A(r),\rw(r),C(r))$ can be extended to some $\rho>r_h$.  We define $\rho$ to be the largest value for which the solution is valid and such that $A(r)<0$ for all $r\in(r_h,\rho)$.

We now define the function
\begin{equation}\label{h}
h(r)=(1-\rw^2)^2-2r^2A{\rw'}^2.
\end{equation}
The global smoothness and asymptotic behavior of $\rw$, $A$ and $C$ will follow from the following:
\begin{Lma}\label{hlemma}
$\lim_{r\nearrow \rho}h(r)$ exists and is finite.
\end{Lma}
\noindent
\textbf{Proof}:  A simple calculation yields 
\begin{equation}\label{h'}
h'(r)=2r{\rw'}^2(\Phi+2A{\rw'}^2-2A).
\end{equation}
where
\begin{equation}\label{Phi}
\Phi(r)=1-A-\frac{(1-\rw^2)^2}{r^2}-\Lambda r^2.
\end{equation}
To simplify notation, we define
\begin{equation}\label{k}
k(r)=\frac{\Phi+2A{\rw'}^2-2A}{r}
\end{equation}
and rewrite equation~(\ref{h'}) as
\begin{equation}\label{heq}
h'=2r^2{\rw'}^2k.
\end{equation}
Another routine calculation yields 
\begin{equation}\label{keq}
r^2k'+2r({\rw'}^2+1)k+2p=0
\end{equation}
where 
\begin{equation}\label{p}
p=1-\frac{2(1-\rw^2)^2}{r^2}-2A{\rw'}^2.
\end{equation}
Clearly, since $r_h>1/\sqrt{\Lambda}$ and $A(r_h)=0$, 
\begin{equation}\label{kneg}
k(r_h)=\frac{\Phi(r_h)}{r_h}<0.
\end{equation}
We claim, furthermore, that $k<0$ for all $r\in (r_h,\rho)$.  For otherwise, let $\tilde r$ be the smallest $r\in(r_h,\rho)$ that satisfies
\begin{equation}\label{tr}
k(\tilde r)=0.
\end{equation}
Equations~(\ref{heq}) and (\ref{h}) imply 
\begin{equation}\label{hcomp}
h(r)<h(r_h)<1\;\mathrm{for}\;\mathrm{all}\;r\in [r_h,\tilde r];
\end{equation}
in particular, $(1-\rw^2(\tilde r))^2<1$ and since $\tilde r>\sqrt{2}$,
\begin{equation}\label{wtilder}
[\frac{2(1-\rw^2)^2}{r^2}]_{r=\tilde r}<1.
\end{equation}
Therefore, 
\begin{equation}\label{p0}
p(\tilde r)>0.
\end{equation}
Now, on one hand, from equations~(\ref{keq}), (\ref{tr}), and (\ref{p0}) it follows that
\begin{equation}\label{k'tilder}
k'(\tilde r)<0.
\end{equation}
On the other hand, because of equation~(\ref{kneg}) and the definition of $\tilde r$, we must have 
\begin{equation}\label{k'tr}
k'(\tilde r)\ge 0.
\end{equation}
Since both equations~(\ref{k'tilder}) and~(\ref{k'tr}) cannot hold, there can be no $\tilde r$ that satisfies equation~(\ref{tr}).  This establishes the claim.  Since $h>0$ for all $r\in(r_h,\rho)$, the Lemma follows. \hfill $\blacksquare$\\ \\
We shall prove that $\rho=\infty$.  Assuming this, we have the following:
\begin{Crly}\label{Aasym}
$A\sim -\frac{\Lambda r^2}{3}$.
\end{Crly}
\noindent
\textbf{Proof}:  Lemma~\ref{hlemma} implies that $\lim_{r\nearrow \infty}[(1-\rw^2)^2/r+2rA{\rw'}^2](r)=0$.  Equation~(\ref{Aeq}) then gives, for any $\epsilon>0$, an $\bar r$ such that 
\begin{equation}\label{A'comp}
|1-\Lambda r^2-(rA)'|<\frac{\epsilon}{r}
\end{equation}
whenever $r>\bar r$.  Integrating equation~(\ref{A'comp}) from $\bar r$ to any $r>\bar r$ implies 
\begin{equation}\label{Acomp}
|1-\frac{\Lambda r^2}{3}+\frac{c}{r}-A|<\frac{\epsilon \ln r}{r}.
\end{equation}
($c$ is a constant of integration.)  The result follows upon taking the limit as $r\nearrow \infty$.  \hfill $\blacksquare$
\begin{Crly}\label{wlimit}
$\lim_{r\nearrow r_h} \rw'(r)=0$ and $\lim_{r\nearrow r_h}\rw(r)$ exists and is finite.
\end{Crly}
\begin{Crly}\label{Climit}
$\lim_{r\nearrow \infty}C(r)$ exists and is finite.
\end{Crly}
\noindent
\textbf{Proof}:  Choosing $C(0)=1$ and integrating equation~(\ref{Ceq})gives
\begin{equation}\label{Csolved}
C(r)=e^{\int_{s=0}^r2{\rw'}^2(s)/s\;rd s}.
\end{equation}
From equation~(\ref{Csolved}) it is clear that $\lim_{r\nearrow \infty}C(r)$ exists although, apriori, it might be infinite.  Lemma~\ref{hlemma} and Corollary~\ref{Aasym} imply ${\rw'}^2\sim r^{-4}$.  Consequently, 
\[\int_0^\infty\frac{2{\rw'}^2}{r}<\infty.\]
\hfill $\blacksquare$\\

Substituting Corollaries~(\ref{Aasym}) and~(\ref{Climit}) into equation~(\ref{metric}) and scaling $t$ by a factor of $[\lim_{r\nearrow \infty}C(r)]^{-1}$  yields a metric that is asymptotic to equation~(\ref{deSitter}).  It is also clear that the Yang Mills field vanishes as $r\nearrow \infty$.\\

It remains to prove the following:
\begin{thm}\label{gbl}
$\rho=\infty$.
\end{thm}
\noindent
\textbf{Proof}:  From Lemma~\ref{hlemma} it follows that $\lp_{r\nearrow \rho}\rw^2(r)<\infty$.  Substituting this into equation~(\ref{Aeq}) gives $\lf_{r\nearrow \rho}A'(r)>-\infty$.  Thus, $\lim_{r\nearrow \rho}A(r)$ exists.
Standard results now imply $\rho<\infty$ only if one of the following holds:  
\begin{list}{(\arabic{prop}):}
{\usecounter{prop}
\lindent}
\item\label{a0}$A(\rho)=0$
\item\label{aneg}$A(\rho)<0$ and $\lp_{r\nearrow \rho}{\rw'}^2(r)=\infty$, or
\item\label{lpw}$\lp_{r\nearrow \rho}{\rw'}^2(r)<\infty$ and $\lp_{r\nearrow \rho}{A'}^2(r)=\infty$.
\end{list}
We now eliminate all three possibilities.\\ \\
\textit{Case~\ref{a0}}.  We first claim that 
\begin{equation}\label{f0}
\lim_{r\nearrow \rho}A{\rw'}^2(r)\;\mathrm{exists\;and\;is\;finite}.
\end{equation}
Clearly, $A{\rw'}^2\le 0$ in the interval $[r_h,\rho]$; i.e., $A{\rw'}^2$ is bounded from above.  It suffices to prove that near $\rho$,  $(A{\rw'}^2)'$ is bounded also from below.

Now, because $\lim_{r\nearrow \rho}A(r)=0$ and $r_h>1/\sqrt{\Lambda}$, there exist $\epsilon>0$ and $\delta>0$ such that $\Phi<-\epsilon$, and consequently,
\begin{equation}\label{phiine}
\Phi+2A{\rw'}^2<-\epsilon,\;\mathrm{whenever}\; r\in (\rho-\delta,\rho).
\end{equation}
Also, Lemma~\ref{hlemma} implies the existence of $M>0$ such that $\rw<M$.  
For arbitrary constant $\beta$, we obtain easily from equations~(\ref{Aeq}) and~(\ref{weq}) the following equation:
\begin{equation}\label{Aw'beta}
r^2(A{\rw'}^\beta)'+r{\rw'}^\beta[(\beta-1)\Phi+2A{\rw'}^2]+\beta\rw{\rw'}^{\beta-1}(1-\rw^2)=0.
\end{equation}
With $\beta=2$, this becomes equation~(\ref{feq})
\begin{equation}\label{feq}
r^2(A{\rw'}^2)'+r{\rw'}^2[\Phi+2A{\rw'}^2]+2\rw\rw'(1-\rw^2)=0.
\end{equation}
From equation~(\ref{feq}) and inequality~(\ref{phiine}) it follows that
\begin{equation}\label{fineq}
(A{\rw'}^2)'\ge\frac{3M(1-M^2)}{\rho^2}
\end{equation}
whenever ${\rw'}^2\le 1$ and $r\in(\rho-\delta,\rho)$.  Also, whenever ${\rw'}^2>1$ and $r\in(\rho-\delta,\rho)$, 
\begin{equation}\label{fineq2}
(A{\rw'}^2)'>\frac{\rw'}{r^2}[r\epsilon \rw'-2M(1-M^2)].
\end{equation}
It is clear from inequalities~(\ref{fineq}) and (\ref{fineq2}) that $(A{\rw'}^2)'$ is bounded from below.  This establishes~(\ref{f0}).

To eliminate this case, we note that on the one hand, $\lim_{r\nearrow \rho}A{\rw'}^2(r)<0$.  Indeed, if $\lim_{r\nearrow \rho}A{\rw'}^2(r)=0$, then equation~(\ref{Aeq}) would give 
\[\lp_{r\nearrow \rho}(rA)'=\lp_{r\nearrow \rho}\Phi(r)\le\lim_{r\nearrow \rho}(1-\Lambda r^2)<0.\]
However, because $rA<0$ throughout the interval $(r_h,\rho)$ and $A(\rho)=0$, we must have $\lp_{r\nearrow \rho}(rA)'(r)\ge 0$.

On the other hand, the assumption $\lim_{r\nearrow \rho}A{\rw'}^2(r)< 0$ leads to a contradiction.  Indeed, since $\lim_{r\nearrow \rho}A(r)=0$, we must have 
\[\lim_{r\nearrow \rho}{\rw'}^2(r)=\infty.\]
The invariance of equations~(\ref{Aeq}) and ~(\ref{weq}) under the transformation $\rw\mapsto -\rw$ allows us to assume, without any loss of generality, that $\lim_{r\nearrow \rho}\rw'(r)=+\infty$.  We next consider equation~(\ref{Aw'beta}) with $\beta=3$ to obtain
\begin{equation}\label{Aw'3}
r^2(A{\rw'}^3)'+r{\rw'}^3(2\Phi+2A{\rw'}^2)+3\rw{\rw'}^2(1-\rw^2)=0.
\end{equation}
Near $\rho$, $\Phi<0$ and thus 
\[2\Phi+2A{\rw'}^2<\lim_{r\nearrow \rho}A{\rw'}^2(r)<-c<0\]
for some $c>0$.
Also, Lemma~\ref{hlemma} implies $\rw$ is bounded.  These facts imply that in equation~(\ref{Aw'3}) the second term on the left dominates the third term on the same side.  It follows that 
\begin{equation}\label{Aw'3ineq}
\lim_{r\nearrow \rho}(A{\rw'}^3)'=+\infty.
\end{equation}
However, $\lim_{r\nearrow \rho}A{\rw'}^2(r)<0$ implies that $\lim_{r\nearrow \rho}A{\rw'}^3=-\infty$.  Therefore 
\[\lf_{r\nearrow \rho}(A{\rw'}^3)'(r)=-\infty.\]
But this contradicts equation~(\ref{Aw'3ineq}).  This eliminates Case~\ref{a0}.\\ \\
\textit{Case~\ref{aneg}}.  Without any loss of generality, we assume that
\[\lp_{r\nearrow \rho}\rw'(r)=+\infty\;\mathrm{and}\; \lim_{r\nearrow \rho}A(r)<0.\]
Thus, there exists a sequence $\{r_n\}\nearrow \rho$ such that
$\lim_{n\rightarrow \infty}(A\rw')(r_n)=-\infty$ and $\lim_{n\rightarrow \infty}(A\rw')'=-\infty$.  Equation~(\ref{Aw'beta}) with $\beta=1$ gives
\begin{equation}\label{veq}
r^2(A\rw')'+2r{\rw'}^2(A\rw')+\rw(1-\rw^2)=0.
\end{equation}
Evaluating equation~(\ref{veq}) at $r_n$ implies $\lim_{n\rightarrow \infty}\rw(r_n)=-\infty$.  We claim that 
\begin{equation}\label{w'i}
\lim_{r\nearrow \rho}\rw(r)=-\infty.
\end{equation}
For if not, then there exist $c>0$ and a sequence $\{s_n\}\nearrow \rho$ such that $\rw(s_n)>-c$, $2s_nA{\rw'}^3(s_n)\nearrow +\infty$, and $(A\rw')'(s_n)>0$.  This in turn implies that, for sufficiently large $n$,
\[[r^2(A\rw')'+2rA{\rw'}^3 +\rw(1-\rw^2)]_{r=s_n}>0,\]
contradicting equation~(\ref{veq}).  This establishes equation~(\ref{w'i}).

Similarly, we assert that 
\[\lf_{r\nearrow \rho}\rw'(r)\ge 0.\]
For otherwise, there exists a sequence $\{t_n\}\nearrow \rho$ such that $A\rw'(t_n)>0$ and $(A\rw')'(t_n)>0$.  Now,
\[[r^2(A\rw')'+2rA{\rw'}^3+\rw(1-\rw^2)]_{r=t_n}>0\]
which also contradicts equation~(\ref{veq}).

Since $\lim_{r\nearrow \rho}\rw(r)=-\infty$ implies $\lf_{r\nearrow \rho}\rw'(r)< 0$, Case~\ref{aneg} is impossible.\\ \\
\textit{Case~\ref{lpw}}.  Equation~(\ref{Aeq}) gives $M>0$ such that
\[(rA)'=1-\frac{(1-\rw^2)^2}{r^2}-\Lambda r^2-2A{\rw'}^2>-M\]
throughout the interval $(r_h,\rho)$.  It follows that $\lim_{r\nearrow \rho}(rA)$, and thus, also $\lim_{r\nearrow \rho}A(r)$ exist and are finite provided $\rho<\infty$. Now, equation~(\ref{Aeq}) written as
\[A'=\frac{1}{r}[1-\frac{(1-\rw^2)^2}{r^2}-\Lambda r^2-2A{\rw'}^2],\]
shows that in the finite interval $[r_h,\rho)$ $A'$ is bounded on both sides.  \hfill $\blacksquare$\\

\bibliographystyle{plain}

\bibliography{bgen}

\end{document}